# Spatial Chern-Simons Interactions and Complex Magnetic Penetration Depth

Yong Tao[†]

College of Economics and Management, Southwest University, Chongqing, China

**Abstract:** This paper examines the Landau-Ginzburg theory in the presence of spatial Chern-Simons interactions, which typically emerge in Weyl semimetals due to domain-wall motion. We demonstrate that the incorporation of a purely spatial Chern-Simons term, which violates parity, into the Landau-Ginzburg free energy leads to a complex magnetic penetration depth. This characteristic indicates that, aside from possessing an effective penetration depth, the magnetic field on the surface of the superconductor experiences periodic spatial oscillations, which significantly deviates from the conventional Meissner effect. In particular, we observe that as the degree of parity breaking, quantified by the strength of the spatial Chern-Simons term, increases, a vortex solution with magnetic field inversion may emerge. With the discovery of superconductivity in certain Weyl semimetals, we anticipate the possibility of experimentally observing this phenomenon in these materials.

[†] Correspondence to: taoyingyong@swu.edu.cn

**Acknowledgement:** This work was supported by the Research project on education and teaching reform in Southwest University (Grant No. 2021JY045)

## 1. Introduction

Recently, the experimental observation of superconductivity in certain Weyl semimetals has attracted significant attention within the physics community [1-6]. It is recognized that Weyl semimetals are a class of systems with chiral fermions [7-9]. In these systems, the chiral anomaly can give rise to electromagnetic interaction terms that violate either parity or time-reversal symmetries [10-13]. These interactions can be represented as Chern-Simons gauge fields [14], which have been utilized to expand the Maxwell equations, resulting in what is known as Maxwell-Chern-Simons electrodynamics [15].

For both (2+1)-dimensional and 3-dimensional systems, the Chern-Simons term is expressed as

$$\mathcal{L}_{CS} \propto \varepsilon^{\mu\nu\rho} A_\mu \partial_\nu A_\rho, \tag{1}$$

where $\varepsilon_{\mu\nu\rho} = \sqrt{g} e_{\mu\nu\rho}$, with $e_{\mu\nu\rho}$ being the totally anti-symmetric pseudo-tensor ($e_{012} = 1$) and $g$ the determinant of the metric tensor $g_{\mu\nu}$.

In (2+1)-dimensional systems, the Chern-Simons term (1) breaks both parity and time-reversal symmetries, a feature linked to intriguing physical phenomena such as the fractional quantum Hall effect [16-18]. Historically, the Chern-Simons term (1) in (2+1)-dimensional systems has been extensively studied. In contrast, for 3-dimensional systems, the Chern-Simons term (1) is purely spatial, violating parity but not time reversal. Spatial Chern-Simons terms are associated with the chiral magnetic effect [19, 20]. It has been demonstrated that the inclusion of spatial Chern-Simons interactions imparts novel characteristics to Maxwell electrodynamics [15]. However, the influence of spatial Chern-Simons interactions on the electrodynamics of superconductors is still under-explored [21, 22]. In fact, spatial Chern-Simons terms naturally emerge in various Weyl systems as a consequence of domain-wall motion [11, 23]. Consequently, with the discovery of superconductivity in certain Weyl semimetals, it is imperative to investigate how spatial Chern-Simons interactions influence the electrodynamics of superconductors. For superconducting systems, it has been proposed that a spatial

Chern-Simons term can be generated from the following gauge-invariant interaction [23]:

$$Re\ \phi^* D_\mu \phi \varepsilon^{\mu\nu\rho} F_{\nu\rho} \rightarrow q\langle\phi\rangle^2 \varepsilon^{\mu\nu\rho} A_\mu \partial_\nu A_\rho, \tag{2}$$

where $D_\mu = \partial_\mu + iqA_\mu$ and $F_{\mu\nu} = \partial_\mu A_\nu - \partial_\nu A_\mu$.

In this paper, we investigate the Landau-Ginzburg theory with the inclusion of a purely spatial Chern-Simons term, which violates parity while preserving time-reversal symmetry. Existing research [22-24] has indicated that this scenario could result in a complex magnetic penetration depth, deviating from the conventional Meissner effect. Here, we further demonstrate that as the degree of parity breaking, quantified by the strength of the spatial Chern-Simons term, increases, a vortex solution featuring magnetic field inversion may emerge. With the discovery of superconductivity in certain Weyl semimetals [1-6], we anticipate the possibility of experimentally observing this phenomenon in these materials. In this paper, we use the natural units $\hbar = c = k_B = 1$, where $\hbar$ denotes the reduced Planck constant, $k_B$ denotes the Boltzmann constant, and $c$ denotes the light speed.

## 2. Spatial Chern-Simons Interactions

Here, we examine the Landau-Ginzburg theory with the inclusion of a purely spatial Chern-Simons term. In this scenario, all the indices in equation (1) are taken to be spatial, where $g_{\mu\nu}$ denotes the Euclidean metric, that is

$$g_{\mu\nu} = g^{\mu\nu} = \mathrm{diag}(1,\quad 1,\quad 1). \tag{3}$$

Based on this setting, for a three-dimensional system, the Landau-Ginzburg free energy density, incorporating a spatial Chern-Simons term, is expressed as:

$$\mathcal{F} = \left[\left(\partial_\mu + iqA_\mu\right)\phi\right]^* \left[\left(\partial^\mu + iqA^\mu\right)\phi\right] + \lambda_2 |\phi|^2 + \lambda_4 |\phi|^4 + \frac{1}{4} F_{\mu\nu} F^{\mu\nu} + \frac{\sigma}{2} \varepsilon^{\mu\nu\rho} A_\mu \partial_\nu A_\rho \tag{4}$$

where $\phi = \left(1/\sqrt{2m}\right)\psi$, with $m$ being the effective mass of the paired electrons, $q$ their effective charge, and $\psi$ the order parameter. Additionally, $\sigma$ is a parameter that quantifies the strength of the spatial Chern-Simons term.

From the free energy density (4), the resulting field equation can be derived as follows:

$$\partial_\mu F^{\mu\nu} = J^\nu \tag{5}$$

with

$$J^\nu = J_{em}^\nu + J_{cs}^\nu, \tag{6}$$

where

$$J_{em}^\nu = -iq(\phi^* \partial^\nu \phi - \partial^\nu \phi^* \phi) + \frac{q^2}{m}|\phi|^2 A^\nu,$$

and

$$J_{cs}^\nu = \frac{\sigma}{2}\varepsilon^{\nu\alpha\beta}F_{\alpha\beta}.$$

Equation (6) implies that the electric current $J^\nu$ comprises two components: $J_{em}^\nu$, recognized as the superconducting current, and $J_{cs}^\nu$, induced by the spatial Chern-Simons term. To elucidate the physical significance of $J_{cs}^\nu$, we introduce the dual tensor:

$$B^\mu = \frac{1}{2}\varepsilon^{\mu\nu\rho}F_{\nu\rho}. \tag{7}$$

Substituting equation (7) into equation (6) we find $J_{cs}^\nu = \sigma B^\nu$. In the context of the Euclidean metric (3), this signifies the chiral magnetic current [19, 20].

For a given superfluid phase stiffness $|\phi|^2$ [25-27], we should have $\phi(x) = |\phi|e^{i\theta(x)}$. Substituting this into equation (6) yields:

$$J_{em}^\nu = \omega_0 A^\nu + \frac{\omega_0}{q}\partial^\nu\theta, \tag{8}$$

where $\omega_0 = 2q^2|\psi|^2$.

Equation (8) implies that the London limit is taken into account. Upon substituting equation (8) into equation (5), we obtain:

$$\partial_\mu F^{\mu\nu} - \frac{\sigma}{2}\varepsilon^{\nu\alpha\beta}F_{\alpha\beta} - \omega_0 A^\nu - \frac{\omega_0}{q}\partial^\nu\theta = 0. \tag{9}$$

## 3. Field equation

To simplify equation (9), we introduce a formula as follows:

$$\varepsilon^{\nu\alpha\beta}\varepsilon_{\nu k\lambda}\partial^\lambda\partial_\alpha A_\beta = \partial_k\partial^\lambda A_\lambda - \partial^\lambda\partial_\lambda A_k. \tag{10}$$

The derivation of equation (10) can be found in Appendix A.

Equation (7) can also be expressed in the following form:

$$B^{\mu} = \varepsilon^{\mu\nu\rho}\partial_{\nu}A_{\rho}. \tag{11}$$

Noting that:

$$\partial_k\partial^{\lambda}A_{\lambda} - \partial^{\lambda}\partial_{\lambda}A_k = \partial^{\lambda}(\partial_k A_{\lambda} - \partial_{\lambda}A_k) = \partial^{\lambda}F_{k\lambda}, \tag{12}$$

by equations (10) and (11), we obtain

$$\partial^{\lambda}F_{k\lambda} = \varepsilon^{\nu\alpha\beta}\varepsilon_{\nu k\lambda}\partial^{\lambda}\partial_{\alpha}A_{\beta} = \varepsilon_{\nu k\lambda}\partial^{\lambda}B^{\nu}. \tag{13}$$

Substituting equations (7) and (13) into equation (9) yields:

$$\varepsilon_{\alpha\nu\beta}\partial^{\beta}B^{\alpha} + \sigma B_{\nu} + \omega_0 A_{\nu} + \frac{\omega_0}{q}\partial_{\nu}\theta = 0. \tag{14}$$

Applying the operator $\varepsilon^{k\lambda\nu}\partial_{\lambda}$ to equation (14), we derive:

$$\varepsilon^{k\lambda\nu}\varepsilon_{\alpha\nu\beta}\partial_{\lambda}\partial^{\beta}B^{\alpha} + \sigma\varepsilon^{k\lambda\nu}\partial_{\lambda}B_{\nu} + \omega_0 B^k + \frac{\omega_0}{q}\varepsilon^{k\lambda\nu}\partial_{\lambda}\partial_{\nu}\theta = 0, \tag{15}$$

where we have used equation (11).

We define:

$$j^k = -\frac{\omega_0}{q}\varepsilon^{k\lambda\nu}\partial_{\lambda}\partial_{\nu}\theta, \tag{16}$$

and thus equation (15) can be rewritten as

$$\varepsilon_{\nu\alpha\beta}\varepsilon^{\nu k\lambda}\partial_{\lambda}\partial^{\alpha}B^{\beta} + \sigma\varepsilon^{k\lambda\nu}\partial_{\lambda}B_{\nu} + \omega_0 B^k = j^k. \tag{17}$$

Equation (16) is interpreted as the vorticity, as referenced in [24]. In the context of the Euclidean metric (3), equation (17) can be expressed in vector form as:

$$\boldsymbol{\nabla} \times \boldsymbol{\nabla} \times \vec{\boldsymbol{B}} + \sigma\boldsymbol{\nabla} \times \vec{\boldsymbol{B}} + \omega_0\vec{\boldsymbol{B}} = \vec{\boldsymbol{j}}. \tag{18}$$

If we consider the superconducting state without vortices, then $\vec{\boldsymbol{j}} = 0$. This corresponds to the situation without a field source:

$$\boldsymbol{\nabla} \times \boldsymbol{\nabla} \times \vec{\boldsymbol{B}} + \sigma\boldsymbol{\nabla} \times \vec{\boldsymbol{B}} + \omega_0\vec{\boldsymbol{B}} = 0. \tag{19}$$

Stalhammar et al. [23] solved equation (19), revealing an oscillating magnetic field with a penetration depth.

Assuming:

$$\boldsymbol{\nabla} \cdot \vec{\boldsymbol{B}} = 0, \tag{20}$$

the vector form (18) simplifies to

$$-\nabla^2\vec{\boldsymbol{B}} + \sigma\boldsymbol{\nabla}\times\vec{\boldsymbol{B}} + \omega_0\vec{\boldsymbol{B}} = \vec{\boldsymbol{j}}. \tag{21}$$

Upon examining equation (21), Vira Shyta et al. [24] also noted an oscillating magnetic field with a penetration depth. Equation (20) implies that magnetic monopoles are artificially excluded, a topic that will be discussed elsewhere.

Equations (19) and (21) are two special cases of equation (18). In this paper, we demonstrate that using the Coulomb gauge, equation (18) can be simplified to a solvable form.

## 4. Complex penetration depth

To simplify equation (18) using the Coulomb gauge, we refer to the tensor form (17) of equation (18). In the context of the Euclidean metric (3), the Coulomb gauge in tensor form is expressed as [28]:

$$\partial^\mu A_\mu = 0. \tag{22}$$

Using the Coulomb gauge (22), equation (17) can be written in a solvable form:

$$\partial^\mu\partial_\mu\partial^\lambda\partial_\lambda B_\nu + (\sigma^2 - 2\omega_0)\partial^\mu\partial_\mu B_\nu + \omega_0^2 B_\nu = \Theta_\nu, \tag{23}$$

where

$$\Theta_\nu = -\sigma\varepsilon_{\nu\rho k}\partial^\rho j^k - \partial^\mu\partial_\mu j_\nu + \omega_0 j_\nu. \tag{24}$$

The derivation for equation (23) is detailed in Appendix B.

In the context of the Euclidean metric (3), equation (23) is equivalent to equation (18). Given this metric, equation (23) can be expressed in vector form as:

$$\nabla^4\vec{\boldsymbol{B}} + (\sigma^2 - 2\omega_0)\nabla^2\vec{\boldsymbol{B}} + \omega_0^2\vec{\boldsymbol{B}} = \vec{\boldsymbol{\Theta}}. \tag{25}$$

Using Swain and Andelman's method [29], the partial differential equation (25) can be exactly solved. Here, we demonstrate that equation (25) may lead to complex magnetic penetration depths. For simplicity, we solve the equation (25) without a field source:

$$\nabla^4\vec{\boldsymbol{B}} + (\sigma^2 - 2\omega_0)\nabla^2\vec{\boldsymbol{B}} + \omega_0^2\vec{\boldsymbol{B}} = 0. \tag{26}$$

Utilizing Swain and Andelman's method [29], we show that equation (26) can be decomposed into two second-order Helmholtz equations. This is achieved by rewriting equation (26) as:

$$(\nabla^2 - \eta_+^2)(\nabla^2 - \eta_-^2)\vec{\boldsymbol{B}} = 0, \tag{27}$$

where

$$\eta_\pm = \frac{\sqrt{4\omega_0 - \sigma^2}}{2} \pm i\frac{\sigma}{2}, \tag{28}$$

and $0 \leq \sigma^2 < 4\omega_0$.

If we define

$$(\nabla^2 - \eta_-^2)\vec{\boldsymbol{B}} = \vec{\boldsymbol{B}}_+, \tag{29}$$

equation (27) can be expressed as:

$$(\nabla^2 - \eta_+^2)\vec{\boldsymbol{B}}_+ = 0. \tag{30}$$

Similarly, by defining

$$(\nabla^2 - \eta_+^2)\vec{\boldsymbol{B}} = \vec{\boldsymbol{B}}_-, \tag{31}$$

equation (27) can be expressed as:

$$(\nabla^2 - \eta_-^2)\vec{\boldsymbol{B}}_- = 0. \tag{32}$$

Using equations (29) to (32), the solution to equation (27) can be written in the form [29]:

$$\vec{\boldsymbol{B}} = c_+\vec{\boldsymbol{B}}_+ + c_-\vec{\boldsymbol{B}}_-, \tag{33}$$

where $c_+$ and $c_-$ are constants.

Thus, solving equation (26) is equivalent to solving two second-order Helmholtz equations as follows:

$$\begin{cases}(\nabla^2 - \eta_+^2)\vec{\boldsymbol{B}} = 0 \\ (\nabla^2 - \eta_-^2)\vec{\boldsymbol{B}} = 0\end{cases}. \tag{34}$$

Equation (34) was also obtained by Stalhammar et al. [23] using a different mathematical approach. Since $\eta_\pm$ are a pair of conjugate complex numbers when $0 \leq \sigma^2 < 4\omega_0$, the real solution of equation (26) can be obtained by solving either of the two Helmholtz equations represented by equation (34). As with the treatment in [23], we solve the Helmholtz equation associated with the root $\eta_+$:

$$(\nabla^2 - \eta_+^2)\vec{\boldsymbol{B}} = 0. \tag{35}$$

In this paper, we focus on exploring the possible vortex solutions of equation (26). To achieve this, we employ cylindrical polar coordinates $(r, \varphi, z)$. Given the magnetic field form $\vec{\boldsymbol{B}} = \big(0,0, B_z(r)\big)$, equation (35) can be transformed into a Bessel equation. Consequently, the real solution to equation (26) can be formally expressed as:

$$B_z(r) = \frac{\Phi_0}{4\pi}[\eta_+^2 K_0(\eta_+ r) + \eta_+^{*2} K_0(\eta_+^* r)], \tag{36}$$

where $\Phi_0 = 1/q$ and $K_0(z)$ represents the modified Bessel function of the second kind.

Since $K_0(z)$ is well-defined for $\mathrm{Re}(z) > 0$, the solution (36) is valid for $0 \leq \sigma^2 < 4\omega_0$. An expression similar to solution (36) has been obtained by [24]. In particular, when $\sigma = 0$, the solution (36) reduces to the London solution:

$$B_z(r) = \frac{\Phi_0 \omega_0}{2\pi} K_0\big(\sqrt{\omega_0} r\big), \tag{37}$$

where the penetration depth is given by $\lambda_p = 1/\sqrt{\omega_0}$.

For $0 < \sigma^2 < 4\omega_0$, $\eta_+$ is a complex number, implying that the penetration depth $\lambda_p = 1/\eta_+$ is also complex. This suggests that the magnetic field $\vec{\boldsymbol{B}}$ on the surface of the superconductor undergoes periodic spatial oscillations, with an effective penetration depth $\lambda_{eff} = 2/\sqrt{4\omega_0 - \sigma^2}$. This behavior deviates from the conventional Meissner effect. Stalhammar et al. [23] have also observed this phenomenon and proposed a potential analogy with optical activity.

Furthermore, for the case where $\sigma^2 - 4\omega_0 \geq 0$, the solution (36) becomes inapplicable since $K_0(z)$ is defined for $\mathrm{Re}(z) > 0$ [24]. In fact, this scenario is quite complex; it may disrupt the positivity of energy [23]. Nonetheless, further analysis indicates that higher-order terms in the Landau-Ginzburg energy functional can restore the positivity of energy [24]. Intriguingly, V. Shyta et al [24] find that this scenario might lead to a novel helimagnetic state [24]. However, the existence of a non-trivial vortex solution in this scenario is questionable since $K_0(z)$ is not well-defined for $\sigma^2 - 4\omega_0 \geq 0$ [24].

In this paper, we examine the possible analytic continuation of the solution (36) for $\sigma^2 - 4\omega_0 \geq 0$. In this regard, we demonstrate that a non-trivial vortex solution with magnetic field inversion may exist in this scenario.

## 5. Vortex solution with magnetic field inversion

When $\sigma^2 - 4\omega_0 \geq 0$, $\eta_+$ becomes a pure imaginary number, rendering $K_0(z)$ in equation (36) undefined directly [24]. To explore the possible analytic continuation of equation (36), we consider the following transformation (refer to page 180 in [30]):

$$\begin{cases} K_0(z) = \frac{i\pi}{2} H_0^{(1)}(iz), & -\pi < arg(z) < \pi/2 \\ K_0(z) = -\frac{i\pi}{2} H_0^{(2)}(-iz), & -\pi/2 < arg(z) < \pi \end{cases}, \tag{38}$$

where $H_0^{(1)}(z)$ and $H_0^{(2)}(z)$ represent the Hankel functions.

Observing that the Bessel function of the second kind can be expressed as (refer to page 175 in [30]):

$$Y_0(z) = \frac{1}{2i}\left[H_0^{(1)}(z) - H_0^{(2)}(z)\right], \tag{39}$$

we derive from equation (38) that:

$$\begin{cases} Y_0(z) = -\frac{1}{\pi}[K_0(-iz) + K_0(iz)] \\ \quad -\pi/2 < arg(z) < \pi/2 \end{cases}. \tag{40}$$

Given that $\eta_+$ is a pure imaginary number for $\sigma^2 - 4\omega_0 \geq 0$ and $arg(r) = 0 \in (-\pi/2, \pi/2)$, we can substitute equation (40) into equation (36) to obtain:

$$B_z(r) = -\frac{\Phi_0}{4}\eta_+^2 Y_0(-i\eta_+ r), \tag{41}$$

Using equations (36) and (41), the real solution of equation (26) can be expressed in a two-stage form:

$$B_z(r) = \begin{cases} \frac{\Phi_0}{2\pi} Re[\eta_+^2 K_0(\eta_+ r)], & 0 \leq \sigma^2 < 4\omega_0 \\ -\frac{\Phi_0}{4}\eta_+^2 Y_0(-i\eta_+ r), & \sigma^2 \geq 4\omega_0 \end{cases}. \tag{42}$$

To investigate the vortex behavior, we consider $r \to 0$. Then, equation (42) gives the circulating current:

$$
J_{\varphi} \sim \begin{cases} \dfrac{\left(\omega_0 - \frac{\sigma^2}{2}\right)\Phi_0}{2\pi r}, & 0 \leq \sigma^2 < 4\omega_0 \\ -\dfrac{\left(\frac{\sigma^2 + \sigma\sqrt{\sigma^2 - 4\omega_0}}{2} - \omega_0\right)\Phi_0}{2\pi r}, & \sigma^2 \geq 4\omega_0 \end{cases}, \tag{43}
$$

where $r > \xi_0$ with $\xi_0$ being the coherence length.

We recognize that $\sigma$ quantifies the strength of the spatial Chern-Simons term, which violates parity. Thus, $\sigma$ serves as a measure of the degree of parity breaking. As shown by equation (43), when $\sigma = 0$, the circulating current yields the London vortex model $J_{\varphi} \sim \omega_0 \Phi_0 / 2\pi r > 0$. In fact, for $\sigma^2 < 2\omega_0$, observing equation (43) we always find $J_{\varphi} > 0$. In this scenario, an increase in $\sigma$ will reduce the current's strength. However, when $\sigma^2 > 2\omega_0$, equation (43) consistently shows $J_{\varphi} < 0$, indicating a magnetic field inversion, as depicted in Figure 1. This suggests that when the degree of parity breaking exceeds a critical threshold, vortices with magnetic field inversion may be observed. The possible diamagnetic effect for the scenario where $\sigma^2 \geq 4\omega_0$ has been discussed in [24, 31]. In contrast, our investigation concentrates on how parity breaking in superconductors induces the emergence of vortices with magnetic field inversion.

## 6. Conclusion

In conclusion, this paper offers an investigation into the Landau-Ginzburg theory in the presence of spatial Chern-Simons interactions, which violates parity. Through meticulous analysis, we have demonstrated that incorporating such interactions can lead to a complex magnetic penetration depth. This characteristic indicates that, aside from possessing an effective penetration depth, the magnetic field on the surface of the superconductor experiences periodic spatial oscillations, which significantly deviates from the conventional Meissner effect. In particular, we have found that as the degree of parity breaking, quantified by the strength of the spatial Chern-Simons term, surpasses a critical threshold, a vortex solution featuring magnetic field inversion may emerge. This suggests that in superconducting materials with significant parity breaking, we may observe vortices with magnetic field inversion. It is known that spatial Chern-Simons terms naturally arise in various Weyl systems due to domain-wall motion. Consequently, with the recent discovery of superconductivity in certain Weyl semimetals, we anticipate that these phenomena will be experimentally observable in these materials.

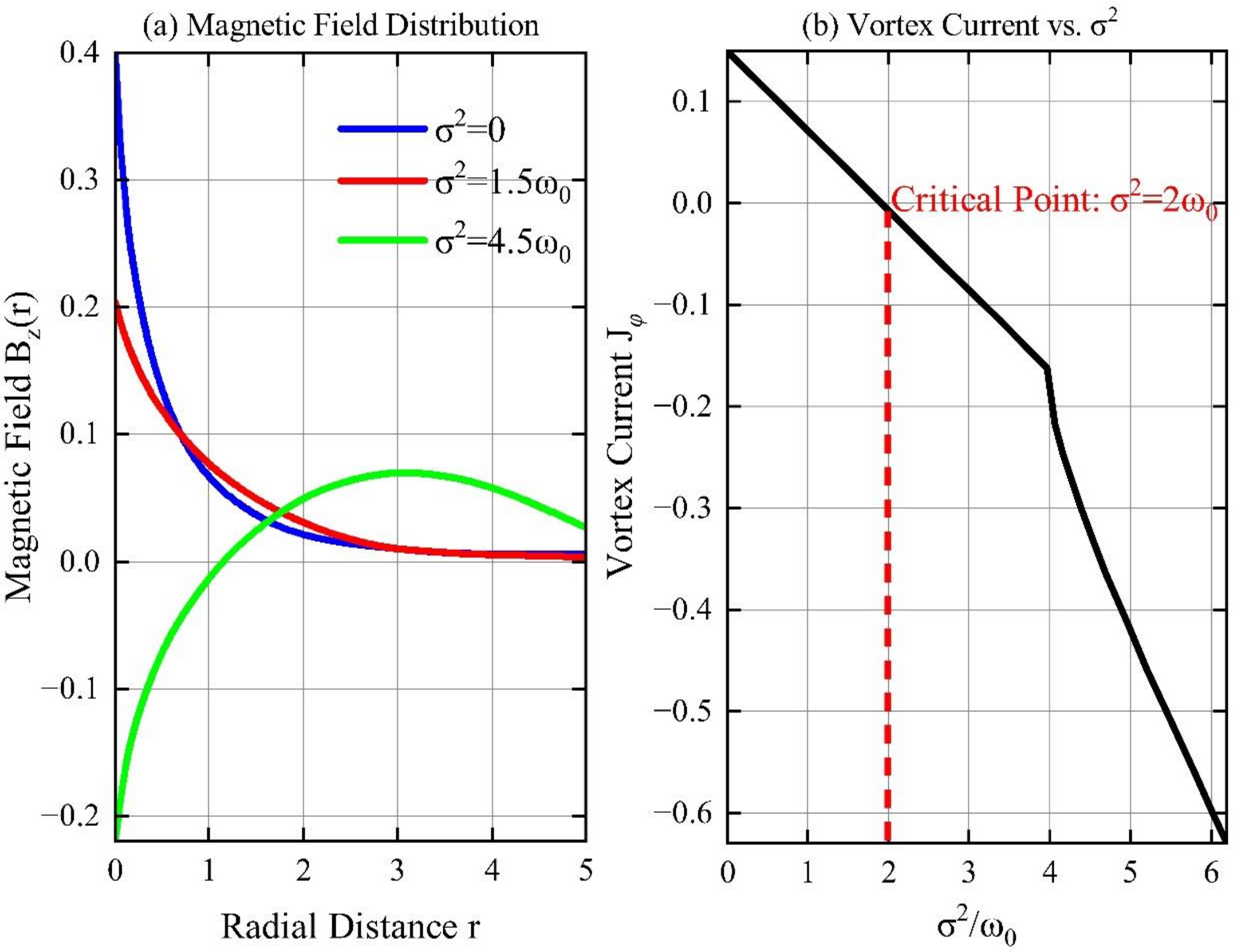

**Figure 1:** The vortex solution with magnetic field inversion

## Appendix A

### The proof for equation (10)

*Proof*. We observe that the left-hand side of equation (10) can be written as the sum of two terms:

$$\varepsilon^{\nu\alpha\beta}\varepsilon_{\nu k\lambda}\partial^{\lambda}\partial_{\alpha}A_{\beta} = \varepsilon^{\nu\lambda\beta}\varepsilon_{\nu k\lambda}\partial^{\lambda}\partial_{\lambda}A_{\beta} + \varepsilon^{\nu\delta\beta}\varepsilon_{\nu k\lambda}\partial^{\lambda}\partial_{\delta}A_{\beta}, \tag{A.1}$$

where $\delta \neq \lambda$.

First, we examine the first term on the right-hand side of equation (A.1), $\varepsilon^{\nu\lambda\beta}\varepsilon_{\nu k\lambda}\partial^{\lambda}\partial_{\lambda}A_{\beta}$.

$$\varepsilon^{\nu\lambda\beta}\varepsilon_{\nu k\lambda}\partial^{\lambda}\partial_{\lambda}A_{\beta} = \frac{1}{\sqrt{g}}e^{\nu\lambda\beta}\sqrt{g}e_{\nu k\lambda}\partial^{\lambda}\partial_{\lambda}A_{\beta} = e^{\nu\lambda\beta}e_{\nu k\lambda}\partial^{\lambda}\partial_{\lambda}A_{\beta} = -e^{\nu\lambda\beta}e_{\nu\lambda k}\partial^{\lambda}\partial_{\lambda}A_{\beta} = -\delta_{k}^{\beta}\partial^{\lambda}\partial_{\lambda}A_{\beta} = -\partial^{\lambda}\partial_{\lambda}A_{k}, \tag{A.2}$$

where $\delta_{k}^{\beta}$ denotes the Kronecker tensor.

Second, we examine the second term on the right-hand side of equation (A.1), $\varepsilon^{\nu\delta\beta}\varepsilon_{\nu k\lambda}\partial^{\lambda}\partial_{\delta}A_{\beta}$. Since $\varepsilon^{\nu\delta\beta}$ and $\varepsilon_{\nu k\lambda}$ are anti-symmetric, $\delta \neq \lambda$ implies $\delta = k$; otherwise, $\delta = \nu$ so that $\varepsilon^{\nu\delta\beta}\varepsilon_{\nu k\lambda}\partial^{\lambda}\partial_{\delta}A_{\beta} = 0$. Therefore, we have

$$\varepsilon^{\nu\delta\beta}\varepsilon_{\nu k\lambda}\partial^{\lambda}\partial_{\delta}A_{\beta} = \varepsilon^{\nu k\beta}\varepsilon_{\nu k\lambda}\partial^{\lambda}\partial_{k}A_{\beta}. \tag{A.3}$$

The right-hand side of equation (A.3) yields:

$$\varepsilon^{\nu k\beta}\varepsilon_{\nu k\lambda}\partial^{\lambda}\partial_{k}A_{\beta} = \frac{1}{\sqrt{g}}e^{\nu k\beta}\sqrt{g}e_{\nu k\lambda}\partial^{\lambda}\partial_{k}A_{\beta} = e^{\nu k\beta}e_{\nu k\lambda}\partial^{\lambda}\partial_{k}A_{\beta} = \delta_{\lambda}^{\beta}\partial^{\lambda}\partial_{k}A_{\beta} = \partial^{\beta}\partial_{k}A_{\beta} = \partial_{k}\partial^{\lambda}A_{\lambda}. \tag{A.4}$$

Substituting equations (A.4) into equation (A.3), we get:

$$\varepsilon^{\nu\delta\beta}\varepsilon_{\nu k\lambda}\partial^{\lambda}\partial_{\delta}A_{\beta} = \partial_{k}\partial^{\lambda}A_{\lambda}. \tag{A.5}$$

Substituting equations (A.2) and (A.5) into equation (A.1), we verify equation (10).

## Appendix B

### The derivation for equation (23)

We first verify that equation (17) can be expressed as

$$\partial^{\mu}\partial_{\mu}B^{k}+\sigma\partial^{\mu}\partial_{\mu}A^{k}-\omega_{0}B^{k}=-j^{k}. \tag{B.1}$$

Utilizing equation (11), the first term of equation (17) can be rewritten as

$$\varepsilon_{\nu\alpha\beta}\varepsilon^{\nu k\lambda}\partial_{\lambda}\partial^{\alpha}B^{\beta}=\varepsilon_{\nu\alpha\beta}\varepsilon^{\nu k\lambda}\partial_{\lambda}\partial^{\alpha}\varepsilon^{\beta\delta\gamma}\partial_{\delta}A_{\gamma}=\varepsilon^{\nu k\lambda}\partial_{\lambda}\varepsilon^{\beta\delta\gamma}\varepsilon_{\nu\alpha\beta}\partial^{\alpha}\partial_{\delta}A_{\gamma}, \tag{B.2}$$

where

$$\varepsilon^{\beta\delta\gamma}\varepsilon_{\nu\alpha\beta}\partial^{\alpha}\partial_{\delta}A_{\gamma}=\partial_{\nu}\partial^{\alpha}A_{\alpha}-\partial^{\alpha}\partial_{\alpha}A_{\nu}. \tag{B.3}$$

Applying the Coulomb gauge (22) and equation (B.3), equation (B.2) simplifies to:

$$\varepsilon_{\nu\alpha\beta}\varepsilon^{\nu k\lambda}\partial_{\lambda}\partial^{\alpha}B^{\beta}=-\varepsilon^{\nu k\lambda}\partial_{\lambda}(\partial^{\alpha}\partial_{\alpha}A_{\nu})=-\partial^{\alpha}\partial_{\alpha}\left(\varepsilon^{\nu k\lambda}\partial_{\lambda}A_{\nu}\right)=-\partial^{\alpha}\partial_{\alpha}B^{k}. \tag{B.4}$$

Using equation (11), the second term of equation (17) can be expressed as

$$\sigma\varepsilon^{k\lambda\nu}\partial_{\lambda}B_{\nu}=\sigma\varepsilon^{k\lambda\nu}\partial_{\lambda}\varepsilon_{\nu\alpha\beta}\partial^{\alpha}A^{\beta}, \tag{B.5}$$

which, by applying equation (10), becomes:

$$\sigma\varepsilon^{k\lambda\nu}\partial_{\lambda}\varepsilon_{\nu\alpha\beta}\partial^{\alpha}A^{\beta}=\sigma\left(\partial^{k}\partial_{\lambda}A^{\lambda}-\partial_{\lambda}\partial^{\lambda}A^{k}\right). \tag{B.6}$$

Substituting the Coulomb gauge (22) and equation (B.6) into equation (B.5) yields:

$$\sigma\varepsilon^{k\lambda\nu}\partial_{\lambda}B_{\nu}=-\sigma\partial^{\lambda}\partial_{\lambda}A^{k}. \tag{B.7}$$

Substituting equations (B.4) and (B.7) into equation (17) verifies equation (B.1).

Now, using Equation (B.1), we derive Equation (23). By applying equation (7), equation (9) can be rewritten as

$$B^{\nu}=\frac{1}{\sigma}\left[\partial_{\mu}(\partial^{\mu}A^{\nu}-\partial^{\nu}A^{\mu})-\omega_{0}A^{\nu}-\frac{\omega_{0}}{q}\partial^{\nu}\theta\right], \tag{B.8}$$

which, using the Coulomb gauge (22), simplifies to:

$$B^{k}=\frac{1}{\sigma}\left(\partial_{\mu}\partial^{\mu}A^{k}-\omega_{0}A^{k}-\frac{\omega_{0}}{q}\partial^{k}\theta\right), \tag{B.9}$$

Substituting equation (B.9) into equation (B.1), we obtain

$$\frac{1}{\sigma}\left[\partial^{\mu}\partial_{\mu}\left(\partial^{\lambda}\partial_{\lambda}A^{k}-\omega_{0}A^{k}-\frac{\omega_{0}}{q}\partial^{k}\theta\right)\right]+\sigma\partial^{\lambda}\partial_{\lambda}A^{k}-\frac{\omega_{0}}{\sigma}\left(\partial_{\mu}\partial^{\mu}A^{k}-\omega_{0}A^{k}-\frac{\omega_{0}}{q}\partial^{k}\theta\right)=-j^{k}. \tag{B.10}$$

Rearranging equation (B.10), we get:

$$\partial^{\mu}\partial_{\mu}\partial^{\lambda}\partial_{\lambda}A^{k}+(\sigma^{2}-2\omega_{0})\partial^{\mu}\partial_{\mu}A^{k}+\omega_{0}^{2}A_{k}=-\sigma j^{k}+\frac{\omega_{0}}{q}\partial^{\mu}\partial_{\mu}\partial^{k}\theta-\frac{\omega_{0}^{2}}{q}\partial^{k}\theta. \tag{B.11}$$

Applying the operator $\varepsilon_{\nu\rho k}\partial^{\rho}$ to equation (B.11), we obtain:

$$\partial^{\mu}\partial_{\mu}\partial^{\lambda}\partial_{\lambda}B_{\nu}+(\sigma^{2}-2\omega_{0})\partial^{\mu}\partial_{\mu}B_{\nu}+\omega_{0}^{2}B_{\nu}=\Theta_{\nu}, \tag{B.12}$$

with

$$\Theta_{\nu}=-\sigma\varepsilon_{\nu\rho k}\partial^{\rho}j^{k}-\partial^{\mu}\partial_{\mu}j_{\nu}+\omega_{0}j_{\nu}, \tag{B.13}$$

where equation (16) has been utilized.